\documentclass[letterpaper]{jpconf}
\usepackage{textcomp}
\usepackage[latin1]{inputenc}
\usepackage{amsfonts}
\usepackage{amssymb}
\usepackage{amsmath}
\usepackage{xspace}
\usepackage{graphicx} 

\newcommand{\FourS}{\ensuremath{\Upsilon(4S)}\xspace}
\newcommand{\OneS}{\ensuremath{\Upsilon(1S)}\xspace}
\newcommand{\TwoS}{\ensuremath{\Upsilon(2S)}\xspace}
\newcommand{\ThreeS}{\ensuremath{\Upsilon(3S)}\xspace}
\newcommand{\nS}{\ensuremath{\Upsilon(nS)}\xspace}
\newcommand{\FiveS}{\ensuremath{\Upsilon(5S)}\xspace}

\newcommand{\invfb}{\ensuremath{\,{\rm fb}^{-1}}\xspace}

\def\mbc{\ensuremath{M_{\rm bc}}\xspace}
\def\deltae{\ensuremath{\Delta E}\xspace}
\def\BR{\ensuremath{\mathcal B}\xspace}

\def\mev{\ensuremath{\hbox{ MeV}}\xspace}
\def\gev{\ensuremath{\hbox{ GeV}}\xspace}
\def\mmev{\ensuremath{\hbox{ MeV}/c^2}\xspace}

\def\bs{\ensuremath{B_s^0}\xspace}
\def\bsst{\ensuremath{B_s^{\ast}}\xspace}

\def\barbsst{\ensuremath{{\bar B_s}^{\ast}}}
\def\bsST{\ensuremath{B_s^{(\ast)}}}
\def\bsSTbsST{\ensuremath{\bsST{\bar B_s}^{(\ast)}}}
\def\KS{\ensuremath{K_S^0}}
\def\pip{\ensuremath{\pi^+}}

\def\kpm{\ensuremath{K^{\pm}}}
\def\ds{\ensuremath{D_s^-}}
\def\dsmp{\ensuremath{D_s^{\mp}}}

\def\bsdspi{\ensuremath{\bs\to\ds\pip}}

\def\bsdsk{\ensuremath{\bs\to\dsmp\kpm}}

\def\r2{\ensuremath{R_2}}

\newcommand{\bfbstodspiun}{[3.67^{+0.35}_{-0.33}({\textrm{stat.}}){}^{+0.43}_{-0.42}({\textrm{syst.}})}
\newcommand{\bfbstodspideux}{\pm0.49(f_s)]\!\times\!10^{-3}}
\newcommand{\bfbstodspi}{\bfbstodspiun\bfbstodspideux}
\newcommand{\fss}{{\left(90.1^{+3.8}_{-4.0}\pm0.2\right)\%}}
\newcommand{\fs}{{\left(7.3^{+3.3}_{-3.0}\pm0.1\right)\%}}
\newcommand{\f}{{\left(2.6^{+2.6}_{-2.5}\right)\%}}

\newcommand{\mbs}{{\left(5364.4\pm1.3\pm0.7\right)\mmev}}
\newcommand{\mbsst}{{\left(5416.4\pm0.4\pm0.5\right)\mmev}}

\newcommand{\bfbstodskun}{[2.4^{+1.2}_{-1.0}({\textrm{stat.}})}
\newcommand{\bfbstodskdeux}{{\pm0.3}({\textrm{syst.}})\pm0.3(f_s)]\!\times\!10^{-4}}
\newcommand{\bfbstodsk}{\bfbstodskun\bfbstodskdeux}

\begin{document}

\title{\boldmath{$\FiveS$} Results at Belle\footnote{Proceedings of a presentation at the Lake Louise Winter Institute 2009 (Alberta, Canada), 16--21 February 2009.}}

\author{Remi Louvot\\{\rm{\it(on behalf of the Belle collaboration)}}}
\address{Laboratoire de Physique des Hautes \'Energies,\\ \'Ecole Polytechnique F\'ed\'erale de Lausanne (EPFL), Lausanne, Switzerland}
\ead{remi.louvot@epfl.ch}
\date{\today}

\begin{abstract}
The data sample recorded with the Belle detector at the KEKB $B$ factory (Tsukuba, Japan) operating at the $\FiveS$ energy provides
interesting and new results about the $\bs$ mesons and the $\FiveS$ resonance.
Recent analyses, based on data samples collected at the $\FiveS$ resonance (23.6 $\invfb$) or near it (7.9 $\invfb$),
are presented with a special focus on the final results on the $\bsdspi$ and $\bsdsk$ decays, and on the intriguing $\FiveS\to\Upsilon(nS)\pi^+\pi^-$ measurements.
\end{abstract}

The Belle experiment \cite{NIMA_479_117}, located at the interaction point of the KEKB asymmetric-energy $e^+e^-$ collider \cite{NIMA_499_1},
was designed for the study of $B$ mesons created by $e^+e^-$ annihilation produced at a center-of-mass (CM) energy corresponding to the mass of the $\FourS$ resonance ($\sqrt s\approx10.58\gev$).
After having recorded an unequaled sample of $\sim800$ millions of $B\bar B$ pairs\footnote{The notation ``$B$'' refers either to a $B^0$ or a $B^+$.
  Moreover, charge-conjugated states are implied everywhere.}, 
the Belle collaboration started to record collisions at higher energies, opening the possibility to study other particles, like the poorly-known $\bs$ meson.
Up to now, 23.6 $\invfb$ of data, containing $\sim2.8$ millions of $\bs$ mesons, have been analyzed at the energy of the $\FiveS$ resonance ($\sqrt s\approx10.87\gev$).

The $\FiveS$ resonance is above the $\bs\bar\bs$ threshold and it was naturally expected that the $\bs$ meson could be studied as well as the $B$ mesons are studied with $\FourS$ data.
The large potential of such $\FiveS$ data was quickly confirmed \cite{PRL_98_052001,PRD_76_012002} with the 2005 engineering run representing 1.86 $\invfb$.
The main advantage with respect to the hadronic colliders is the possibility of measurements of absolute branching fractions.
However, the abundance of $\bs$ mesons in $\FiveS$ hadronic events has to be precisely determined.
Above the $e^+e^-\to u\bar u, d\bar d, s\bar s, c\bar c$ continuum events, the $e^+e^-\to b\bar b$ process can produce different kinds of final states:
seven with a pair of non-strange $B$ mesons ($B^{\ast}\bar B^{\ast}$, $B^{\ast}\bar B$, $B\bar B$, $B^{\ast}\bar B^{\ast}\pi$, $B^{\ast}\bar B\pi$, $B\bar B\pi$ and $B\bar B\pi\pi$)
and three with a pair of $\bs$ mesons ($\bsst\barbsst$, $\bsst\bar\bs$ and $\bs\bar\bs$) since the $B^{\ast}$ and $\bsst$ mesons always decay by emission of a photon.
The total $e^+e^-\to b\bar b$ cross section at the $\FiveS$ energy was measured to be $\sigma_{b\bar b}^{\FiveS}=(302\pm14)$~pb \cite{PRL_98_052001,PRD_75_012002}
and the fraction of $\bs$ events to be $f_s=\sigma(e^+e^-\to\bsSTbsST)/\sigma_{b\bar b}^{\FiveS}=(19.3\pm2.9)$~\% \cite{PLB_667_1}.
The dominant $\bs$ production mode, $b\bar b\to\bsst\barbsst$, represents approximately 90\% \cite{PRL_102_021801} of the $b\bar b\to\bsSTbsST$ events.
Published results on the $\bsdspi$ and $\bsdsk$ modes \cite{PRL_102_021801}
and on the electromagnetic penguin decays $\bs\to\phi\gamma$ and $\bs\to\gamma\gamma$ \cite{PRL_100_121801} are presented.
Preliminary results about the $\bs\to J/\psi\,\KS$ and $\bs\to J/\psi\,\phi$ modes \cite{confICHEP08} and the semi-leptonic $\bs$ decays~\cite{hepex_0710_2548} are also described.
In addition, recent results on bottomonium production are reported \cite{PRL_100_112001},
including preliminary measurements obtained with the data from the energy scan performed near the $\FiveS$ resonance \cite{hepex_0808_2445}.



For the exclusive modes, the $\bs$ candidates are fully reconstructed from the final-state particles.
The signal is analyzed with the successful method developed at the $B$ factories.
From the reconstructed four-momentum in the CM $(E_{\bs}^{\ast},p_{\bs}^{\ast})$, two variables are formed:
the energy difference $\deltae=E_{\bs}^{\ast}-\sqrt s/2$ and the beam-constrained mass $\mbc=\sqrt{s/4-p_{\bs}^{\ast2}}$.
The signal yields are extracted from a two-dimensional fit performed of the distribution of these two variables.
As the $\bs$ can be produced via three kinematically-different $\FiveS$ decays\footnote{In this context, the notation ``$\FiveS$'' stands for any produced $b\bar b$ pair,
including non-resonant $b\bar b$ continuum since it is not distinguishable from the resonant $\FiveS$ state.},
we expect three signal regions in the $(\mbc,\deltae)$ plane.
The location of these regions (two observables per region) can be related to the $\bsst$ and $\bs$ masses,
providing a measurement of these two interesting physical parameters.

\begin{figure}[!h]
\centering

\begin{minipage}[t]{0.5\linewidth}
\includegraphics[width=\linewidth]{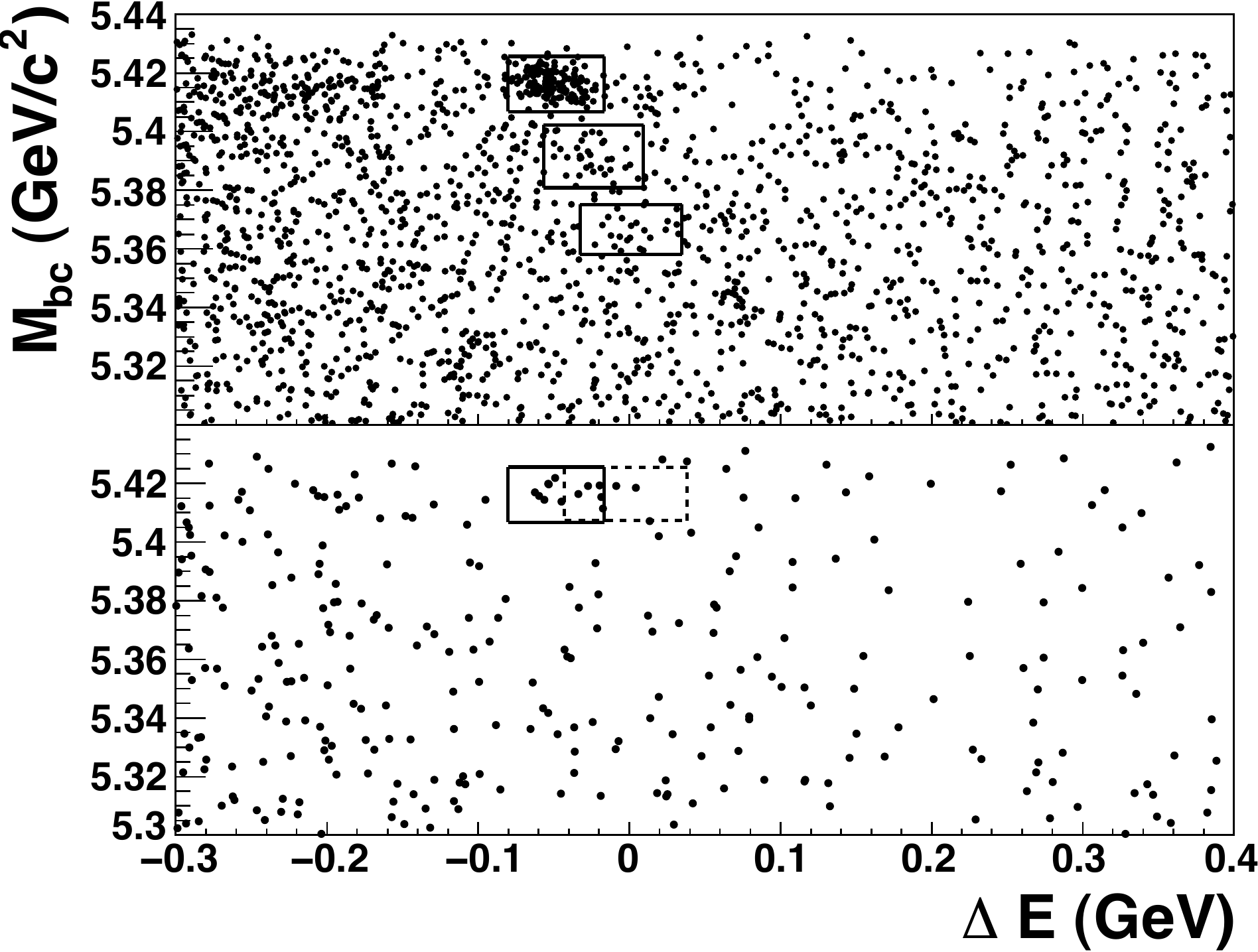}
\end{minipage}~~
\begin{minipage}[b]{.45\linewidth}
\caption{\label{fig:bsdspi:scplot}
  $\bsdspi$ (top) and $\bsdsk$ (bottom) candidates selected in 23.6 $\invfb$ and represented in the $(\mbc,\deltae)$ plane.
  On the top plot, the three boxes represent the 2.5$\sigma$ regions where $\bs$ candidates from $\FiveS\to\bsst\barbsst$, $\bsst\bar\bs$, $\bs\bar\bs$ modes (from left to right) are expected.
  On the bottom plot, the solid box represents the signal region, while the dashed box shows the location of $\bsdspi$ decays when the pion is misidentified as a kaon;
  both boxes are for the $\FiveS\to\bsst\barbsst$ mode only.}
\end{minipage}

\end{figure}

The flavour-specific $\bsdspi$ mode proceeds dominantly via a tree amplitude.
Its Cabibbo-suppressed counterpart, $\bsdsk$, is not flavour-specific and is therefore interesting for $CP$-violation studies.
The dominant $\bsdspi$ mode is a good candidate for a normalization channel at hadron colliders thanks to its clean signature (four charged tracks) and its large branching fraction.
The $\ds$ mesons are reconstructed via three channels: $\ds\to\phi(\to K^+K^-)\pi^-$, $\ds\to K^{\ast0}(\to K^+\pi^-)K^-$ and $\ds\to\KS(\to\pi^+\pi^-)K^-$.
The selected candidates are shown in Fig.~\ref{fig:bsdspi:scplot}.

From the fitted $\bsdspi$ yields and peak positions in the three signal regions, we measure six parameters:
$\BR(\bsdspi)=\bfbstodspi$, $f_{\bsst\bar\bsst}=N_{\bsst\bar\bsst}/N_{\bsSTbsST}=\fss$, $f_{\bsst\bar\bs}=N_{\bsst\bar\bs}/N_{\bsSTbsST}=\fs$,
$f_{\bs\bar\bs}=N_{\bs\bar\bs}/N_{\bsSTbsST}=\f$, $m_{\bs}=\mbs$ and $m_{\bsst}=\mbsst$.
For the $\bsdsk$ candidates, we fit a signal only in the $\bsst\bar\bsst$ region as 90\% of the events are concentrated there.
A 3.5$\sigma$ evidence with $6.7^{+3.4}_{-2.7}$ events is obtained, leading to the branching fraction $\BR(\bsdsk)=\bfbstodsk$.

\begin{figure}[!h]
\centering
\begin{minipage}{0.4\linewidth}
\includegraphics[width=\linewidth,height=3.5cm]{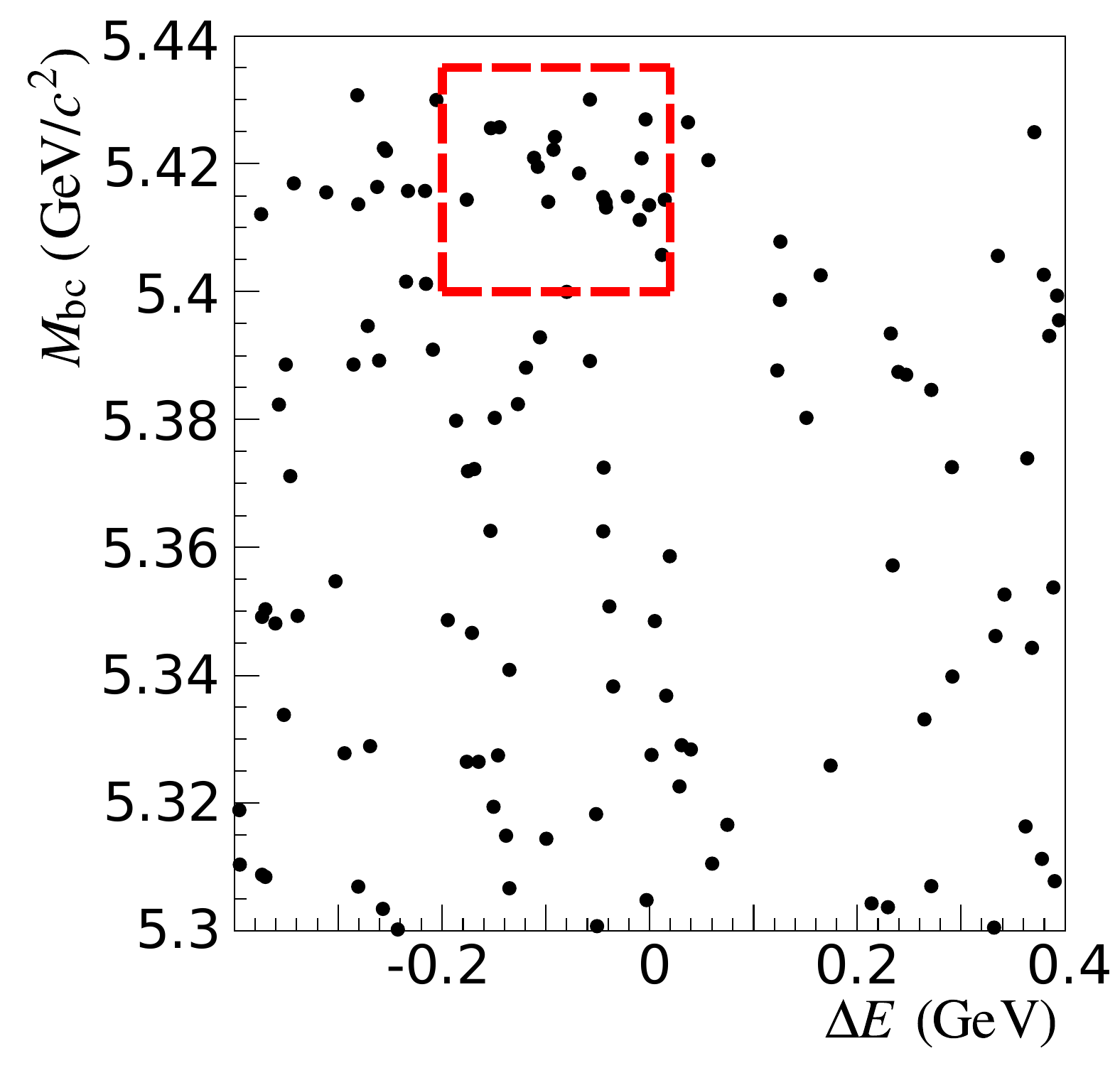}
\caption{\label{fig:bsphigamma}Selected $\bs\to\phi\gamma$ candidates. The box is the 2.5$\sigma$ region where the signal is expected for the $\FiveS\to\bsst\barbsst$ mode.}
\end{minipage}
~~~
\begin{minipage}{0.4\linewidth}
\includegraphics[width=\linewidth,height=3.5cm]{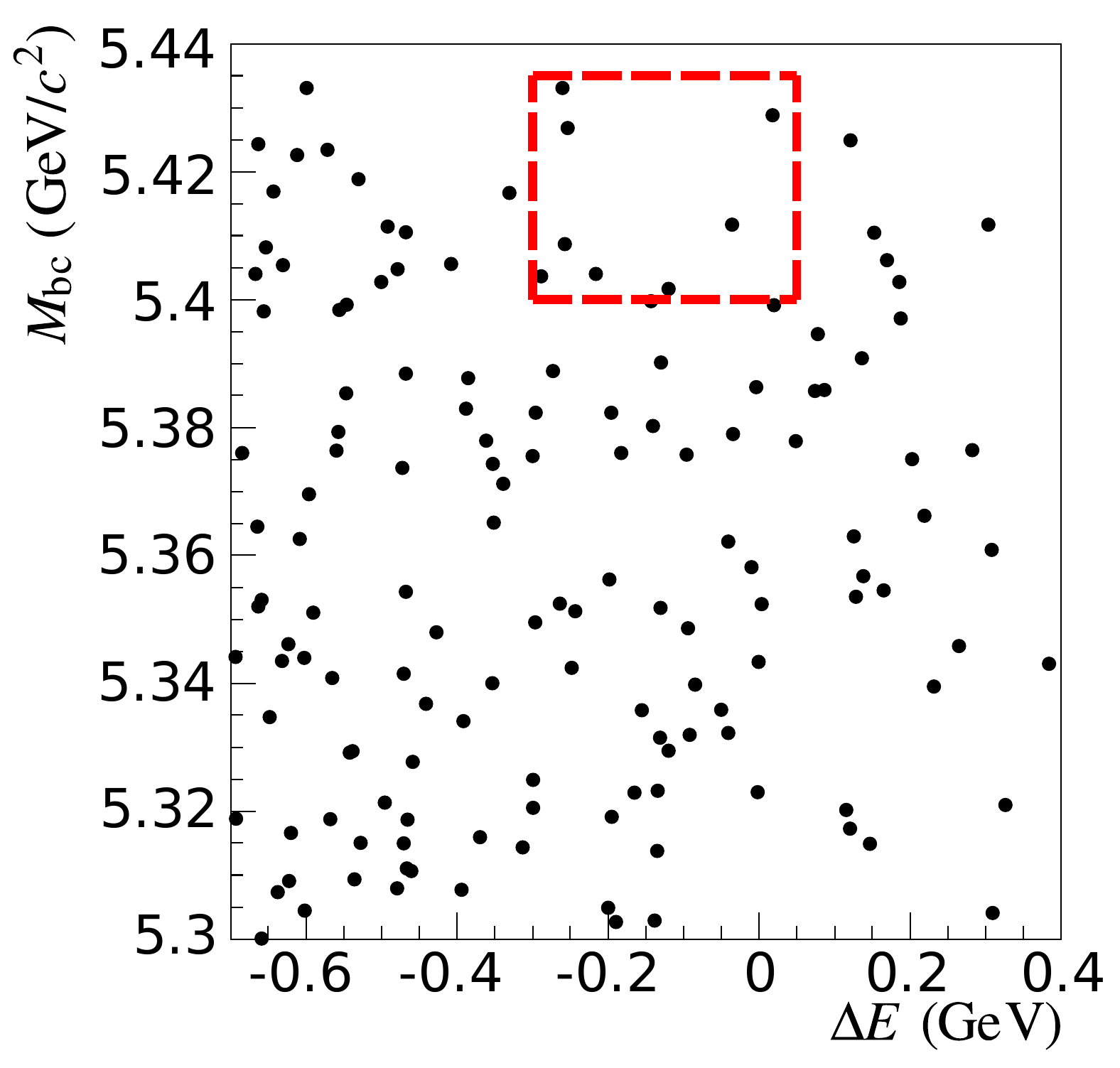}
\caption{\label{fig:bsgammagamma}Selected $\bs\to\gamma\gamma$ candidates. The box is the 2.5$\sigma$ region where the signal is expected for the $\FiveS\to\bsst\barbsst$ mode.}
\end{minipage}
\end{figure}

The dominant process leading to the $\bs\to\phi\gamma$ and $\bs\to\gamma\gamma$ decays is an electromagnetic radiative penguin diagram.
The $b\to s\gamma$ transitions are an important test for the standard model (SM).
We obtain the first observation ($5.5\sigma$) with $18^{+6}_{-5}$ events (Fig.~\ref{fig:bsphigamma}),
leading to the branching fraction $\BR(\bs\to\phi\gamma)=(57^{+18}_{-15}{}^{+12}_{-11})\times10^{-6}$.
This is the first observation of a radiative $\bs$ decay.
We don't have enough statistics to see a significant $\bs\to\gamma\gamma$ excess (Fig.~\ref{fig:bsgammagamma}).
We set, at 90\% C.L., an upper limit $\BR(\bs\to\gamma\gamma)<8.7\times10^{-6}$, which is six times more stringent than the previous limit
and only one order of magnitude larger than the SM prediction.

\begin{figure}[!h]
\centering

\begin{minipage}[t]{0.6\linewidth}
 \includegraphics[width=\linewidth]{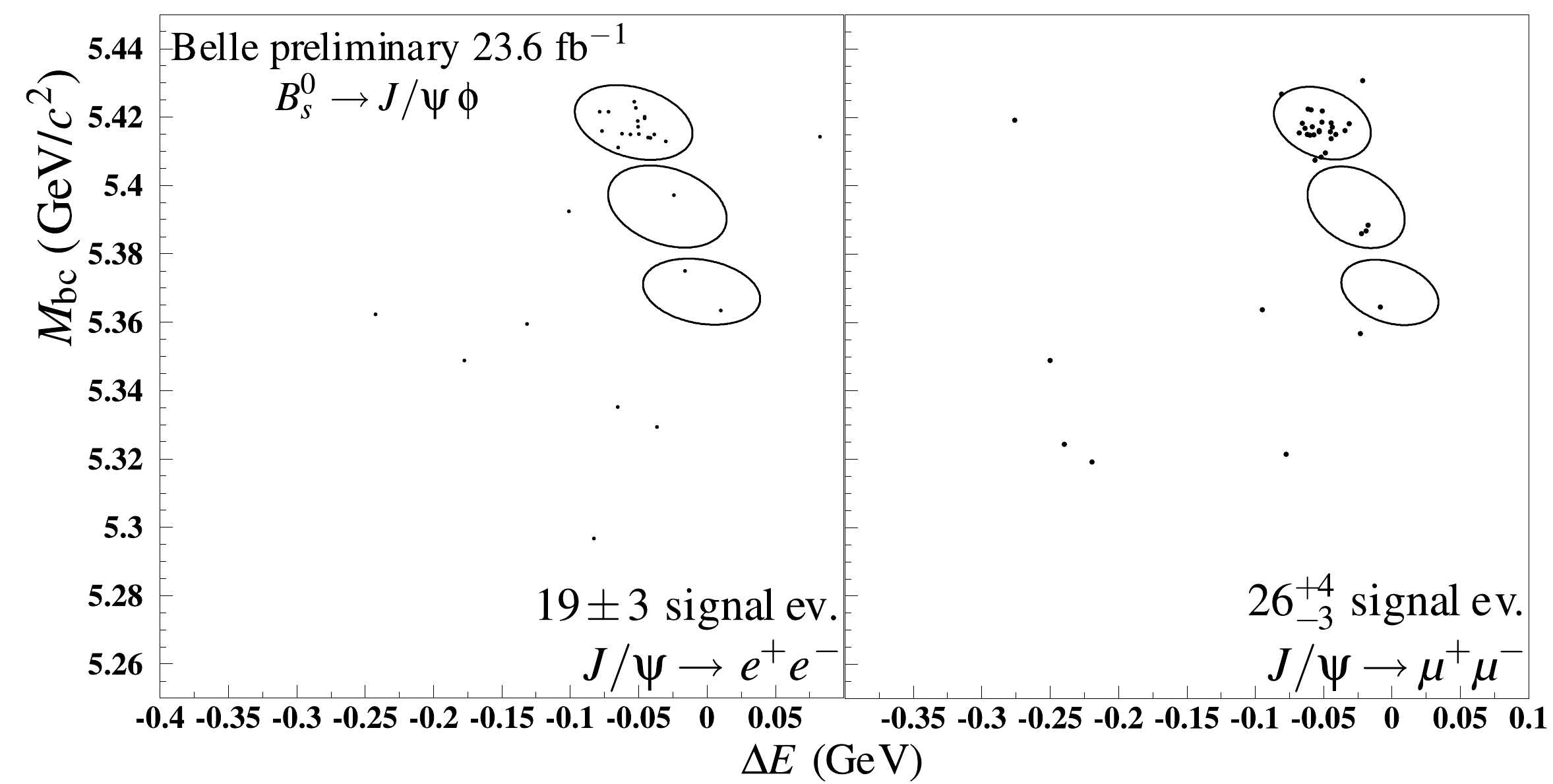}
\end{minipage}~~
\begin{minipage}[b]{.35\linewidth}
\caption{\label{fig:jpsiphi}Selected $\bs\to J/\phi\,\phi$ candidates. The left (right) plot presents candidates with the $J/\phi$ decaying to electrons (muons).
The elliptic regions have the same meaning as the boxes in the top plot of Fig.~\ref{fig:bsdspi:scplot}.
}
\end{minipage}
\end{figure}

A search for the $\bs\to J/\psi\,\phi$ decay (and the Cabibbo-suppressed $\bs\to J/\psi\,\KS$ decay) has been performed (Fig.~\ref{fig:jpsiphi}).
The leading contribution comes from the colour-suppressed $b\to c\bar cs$ ($c\bar cd$) spectator diagram, but a penguin loop may also contribute.
While no significant signal is seen for the $\bs\to J/\phi\,\KS$ mode,
the observation of $\sim45$ events for the $\bs\to J/\psi\,\phi$ mode leads to
the first absolute measurement of the branching fraction $\BR(\bs\to J/\psi\,\phi)=(1.15^{+0.28}_{-0.30})\times10^{-3}$.

The inclusive semi-leptonic branching fractions of the $\bs$ meson have been measured thanks to its fast particle-antiparticle oscillation:
by requiring events with a fully reconstructed $\ds\to\phi(\to K^+K^-)\pi^-$ and a fast lepton with the same sign,
only events with a $\bs\bar\bs$ pair are selected, 
After a fit of the lepton CM momentum to disentangle primary and secondary leptons,
the results obtained are $\BR(\bs\to X^+\,e^-\,\nu)=(10.9\pm1.0\pm0.9)\%$ and $\BR(\bs\to X^+\,\mu^-\,\nu)=(9.2\pm1.0\pm0.8)\%$.
The average is $\BR(\bs\to X^+\,l^-\nu)=(10.2\pm0.8\pm0.9)\%$, in good agreement with $\BR(B^0\to X^+\,l^-\nu)$ \cite{PLB_667_1} which is expected to be very similar.

\begin{figure}[!h]
\centering

\begin{minipage}[t]{0.5\linewidth}
\includegraphics[width=\linewidth]{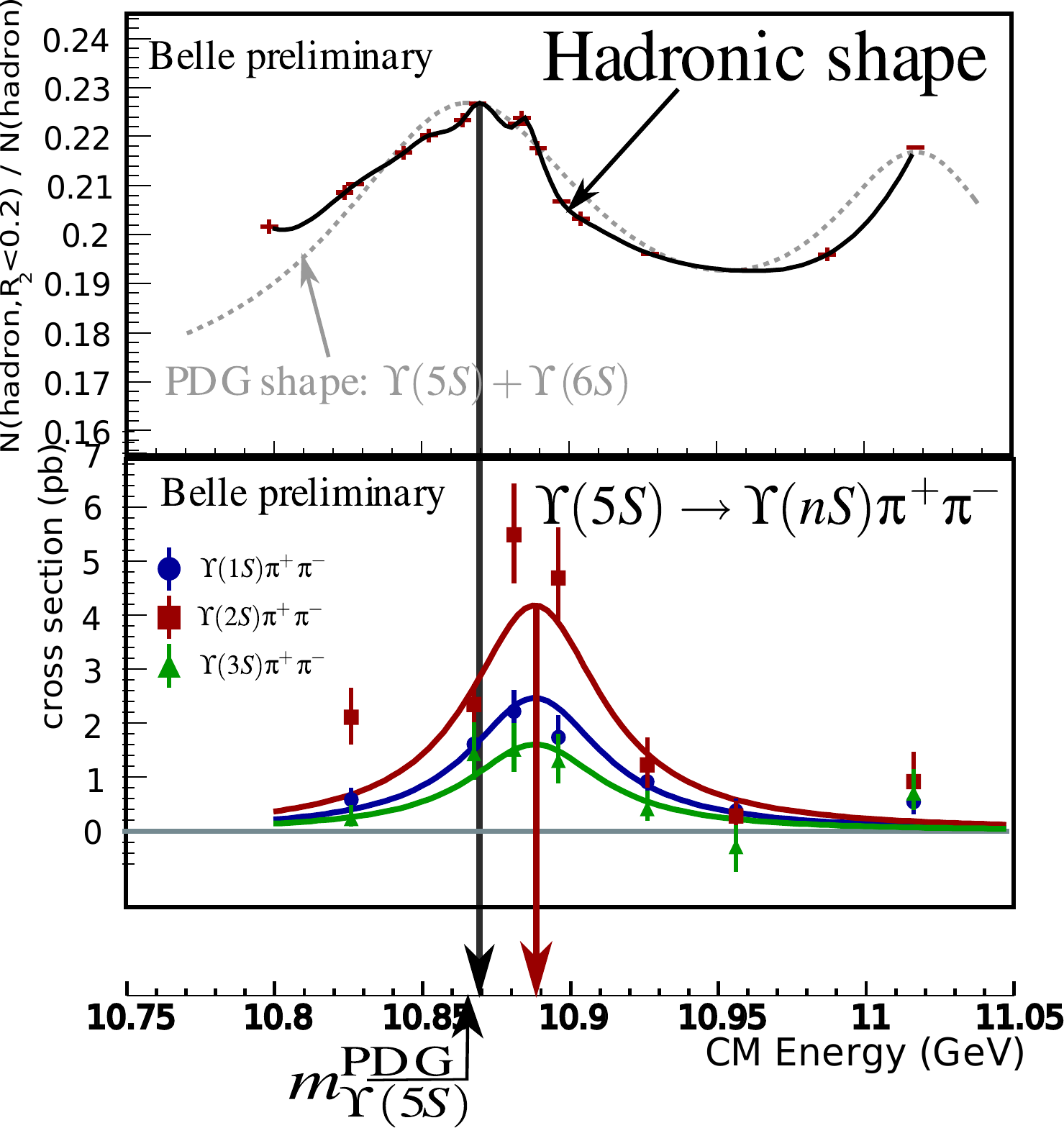}
\end{minipage}
~~~
\begin{minipage}[b]{0.4\linewidth}
\caption{\label{fig:escan} Hadronic production (top) and $\FiveS\to\nS\pi^+\pi^-$ cross section (bottom) as a function of $\sqrt s$.
On the top plot, the grey dashed line is the shape of the $\FiveS$ and $\Upsilon(6S)$ as present in Ref.~\cite{PLB_667_1}.
On the bottom plot, the three sets of points, representing $\FiveS\to\OneS\pi^+\pi^-$ (circle, blue),
$\FiveS\to\TwoS\pi^+\pi^-$ (square, red) and $\FiveS\to\ThreeS\pi^+\pi^-$ (triangle, green) data,
are fitted with the same Breit-Wigner shape.
}
\end{minipage}

\end{figure}

Not only $\bs$ mesons can be studied with $\FiveS$ data.
A search for bottomonium production ($\nS$, $n=1,2,3$), based on the full reconstruction of $\FiveS\to\nS(\to \mu^+\mu^-)\pi^+\pi^-$ and $\FiveS\to\nS(\to\mu^+\mu^-)K^+K^-$ decays,
showed rates about two order of magnitude larger than expected from $\FourS$ rates. 
A six-point energy scan near the $\FiveS$ resonance has been performed to study the variation of these rates as function of $\sqrt s$.
On Fig.~\ref{fig:escan}, a clear difference between the inclusive hadronic line shape ($m=10865\pm8\,\mmev$, $\Gamma=110\pm13\,\mev$)
and the exclusive bottomonium-production line shape ($m=10889.6\pm2.3\,\mmev$, $\Gamma=54.7^{+8.9}_{-7.6}\,\mev$) can be seen. 
While the former is compatible with previous CLEO and CUSB measurements \cite{PRL_54_381,PRL_54_377}, the latter is shifted by $+(24.6\pm8.3)\mmev$ and twice narrower.
An interpretation could be a new $Y_b$ state with small production cross section and large branching fraction to the $\Upsilon(nS)\pi\pi$ final state.
However the Babar collaboration has recently measured \cite{PRL_102_012001} an inclusive line shape with a width twice smaller than Belle.
The result of an exclusive analysis by Babar would be very welcome to help clarifying the situation.

To conclude, all these studies demonstrate the great potential of the Belle dataset recorded at $\FiveS$ energy.
The sensitivity obtained for several $\bs$ modes allows many interesting measurements, from $\bs$ physical parameters to searches for new physics.
Besides that, the intriguing measurements of the $\FiveS\to\nS h^+h^-$ channels open a new interest in bottomonium spectroscopy.
So far, the full Belle sample has reached 65 $\invfb$, and the KEKB collider will continue delivering collisions at the $\FiveS$ energy during 2009.
Of course, many more interesting results are expected with the full Belle $\FiveS$ dataset.

~

~








\providecommand{\newblock}{}


\end{document}